\documentclass[a4paper,twoside]{article}

\usepackage{epsfig}
\usepackage{subcaption}
\usepackage{calc}
\usepackage{amssymb}
\usepackage{amstext}
\usepackage{amsmath}
\usepackage{amsthm}
\usepackage{multicol}
\usepackage{multirow}
\usepackage{url}
\usepackage{hyperref}

\usepackage{pslatex}
\usepackage{apalike}
\usepackage{algorithm2e}
\usepackage[bottom]{footmisc}
\usepackage{mhchem}

\usepackage[table]{xcolor} 
\usepackage{booktabs}      
\usepackage[dvipsnames, table]{xcolor}
\definecolor{firstColor}{HTML}{63C07A}
\definecolor{secondColor}{HTML}{BDDC8E}
\definecolor{thirdColor}{HTML}{EBE7A0}

\usepackage{SCITEPRESS}     

\begin{document}

\title{Tracking Urban Atmospheric Pollutants using Sentinel‑5P Satellite Data}


\author{\authorname{Alice Gomez-Cantos\sup{1}\orcidAuthor{0000-0002-2786-2677}, Henry O. Velesaca\sup{2}\orcidAuthor{0000-0003-0266-2465}}
\affiliation{\sup{1}Facultad de Ciencias Naturales y Matemáticas, Escuela Superior Politécnica del Litoral, ESPOL, Campus Gustavo Galindo, Km. 30.5 Vía Perimetral, Guayaquil, 090902, Ecuador}
\affiliation{\sup{2}Software Engineering Department, Research Center for Information and Communication Technologies (CITIC-UGR), University of Granada, 18071, Granada, Spain}
\email{\{alivagom, hvelesac\}@espol.edu.ec}
}



\abstract{Urban nitrogen dioxide ($NO_2$) is a key indicator of combustion-related air pollution and exhibits strong spatial and temporal variability in cities. This study presents a satellite-based framework for tracking urban $NO_2$ pollution using tropospheric column observations from Sentinel-5P/TROPOMI over Guayas Province, Ecuador. Rather than estimating surface concentrations, the methodology emphasizes robust distributional metrics, including the median and upper-tail percentiles ($P_{90}$, $P_{95}$, and $P_{99}$), to characterize background conditions and localized pollution extremes at the canton scale. Multi-year satellite observations are aggregated annually and analyzed using unsupervised K-means clustering to identify characteristic pollution regimes without predefined thresholds. Results show that highly urbanized cantons consistently exhibit elevated extreme $NO_2$ values and greater variability, while less urbanized areas display lower and more homogeneous patterns. The proposed approach provides an interpretable and scalable tool for urban air-quality assessment in data-scarce regions using satellite observations alone. The implementation is publicly available on GitHub \url{https://hvelesaca.github.io/sentinel-5P-clustering/}.
}

\onecolumn \maketitle \normalsize \setcounter{footnote}{0} \vfill

\section{\uppercase{Introduction}}
\label{sec:introduction}
Urban nitrogen dioxide (NO$_2$) is a key air-quality pollutant because it is tightly coupled to high-temperature combustion sources (e.g., road traffic, power generation, and industrial activity) and therefore exhibits strong spatial heterogeneity over cities \cite{fenger1999urban,Judd2020AMT}. As a short-lived species, NO$_2$ responds rapidly to changes in emissions and meteorological transport, making satellite observations particularly valuable for tracking urban-scale variability \cite{andreae2019emission}. In addition to its direct health relevance, NO$_2$ plays an important role in atmospheric chemistry: in the presence of sunlight and volatile organic compounds, NO$_2$ contributes to photochemical ozone production and to broader processes affecting air-quality and oxidation capacity \cite{lorente2021methane,mejia2023sentinel}.

Satellite remote sensing complements in situ monitoring by providing spatially continuous coverage. The Sentinel-5 Precursor mission (Sentinel-5P) carrying the TROPOspheric Monitoring Instrument (TROPOMI) provides near-daily global observations of NO$_2$ at kilometer-scale resolution, enabling the identification of emission hotspots and the characterization of urban pollution gradients \cite{tian2022investigating,fiore2015air}. However, the operational TROPOMI NO$_2$ product is a column quantity rather than a direct near-surface concentration, so urban interpretation is best framed in terms of spatiotemporal contrasts, anomalies, and trends in retrieved columns \cite{davybida2023air,sha2021validation}.

Robust use of TROPOMI NO$_2$ for city-scale analyses requires careful spatiotemporal processing because satellite observations are high-dimensional, noisy, and irregularly sampled across space and time \cite{Verhoelst2021AMT,Douros2023GMD,Judd2020AMT}. In this context, machine learning (ML) provides a practical framework to transform multi-year NO$_2$ columns into interpretable urban indicators by (i) learning dominant spatiotemporal structures, (ii) separating persistent patterns from short-term variability, and (iii) grouping areas with similar pollution behavior using unsupervised methods (e.g., clustering \cite{velesaca2024analysis}). Building on these foundations, this work integrates machine learning with Sentinel-5P/TROPOMI NO$_2$ to track urban atmospheric pollution patterns and to derive robust summary statistics that enable comparison of temporal changes and spatial heterogeneity across the study region.

The manuscript is organized as follows. Section \ref{sec:back} presents studies related to tracking urban atmospheric pollutants using satellite data. Section \ref{sec:method} presents the proposed methodology. Experimental results and comparisons are given in Section \ref{sec:ExpRes}. Finally, conclusions are presented in Section \ref{sec:conclusion}.

\section{\uppercase{BACKGROUND}}
\label{sec:back}

Recent studies show an increasing use of Sentinel-5P/TROPOMI data for urban air-quality assessment, with diverse analytical strategies and complementary strengths. In data-scarce contexts, Mej\'{\i}a et al.~\cite{mejia2023sentinel} demonstrate how spatial interpolation of Sentinel-5P NO$_2$ can improve intra-urban interpretability in Guayaquil, highlighting empirical Bayesian kriging as an effective approach when ground monitoring is limited. However, interpolation-based methods remain constrained by the native satellite resolution and cannot generate new sub-pixel information.

At the city scale, Shah et al.~\cite{shah2024air} illustrate the operational feasibility of satellite-based urban air-quality monitoring through spatiotemporal analysis over Pune City. Their work shows that repeated TROPOMI observations can be transformed into interpretable urban indicators, although results remain sensitive to processing choices and to the column-based nature of the satellite product.

More recently, Prajesh et al.~\cite{prajesh2025satellite} propose a signal-isolation framework that exploits seasonal structure and reference-region contrasts to extract differential pollution signals from Sentinel-5P time series. While this approach strengthens attribution-like interpretations, it also highlights challenges related to non-stationarity and evolving background conditions in atmospheric columns.

Overall, existing literature suggests that Sentinel-5P/TROPOMI enables robust characterization of urban NO$_2$ patterns when analyses focus on spatiotemporal contrasts and distributional behavior rather than absolute surface concentrations. These insights motivate the present work, which emphasizes robust statistical metrics and unsupervised learning to track urban pollution dynamics under the constraints of column-based satellite observations.

\section{\uppercase{Material \& Methods}}
\label{sec:method}
The proposed methodology integrates satellite-based remote sensing data from Sentinel-5P/TROPOMI with robust statistical analysis to characterize the spatiotemporal behavior of urban $NO_2$ pollution, taking Guayas Province in Ecuador as a case study. Rather than attempting to infer surface-level concentrations, the methodology focuses on relative contrasts, temporal evolution, and upper-tail behavior of tropospheric $NO_2$ columns at the canton scale. The workflow is designed to be reproducible, data-driven, and suitable for regions with limited ground-based air-quality monitoring.

\begin{figure*}[!h]
    \centering
    \includegraphics[width=0.75\textwidth]{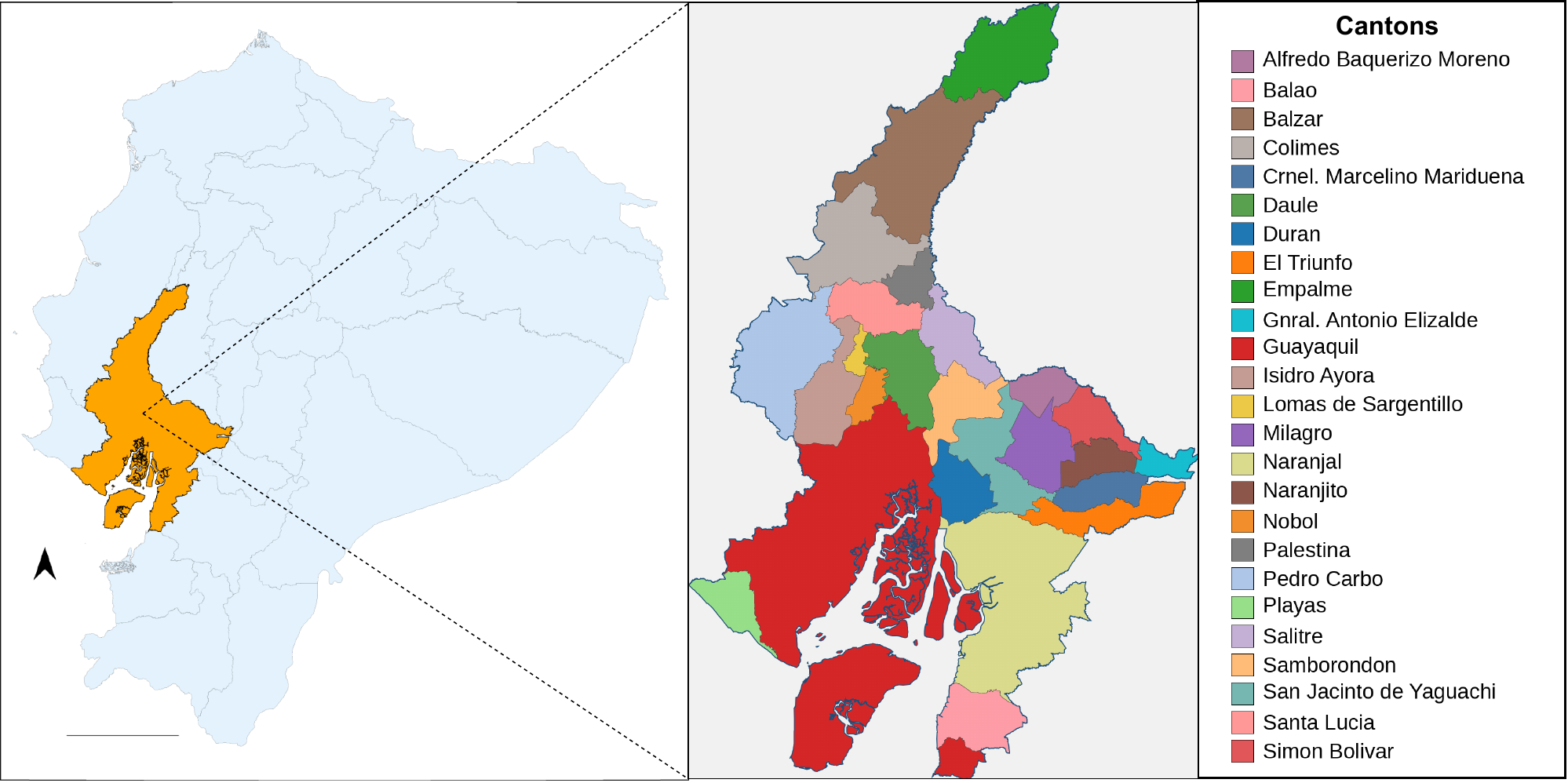}
    \caption{Map of Guayas Province in Ecuador, illustrating the spatial distribution of cantons across the province.}
    \label{fig:map_guayas}
\end{figure*}

Figure \ref{fig:map_guayas} illustrates the geographical context of the study area, highlighting the spatial distribution of cantons within Guayas Province. The overall methodology consists of four main stages: (i) data acquisition and preprocessing, (ii) spatial aggregation at the canton level, (iii) temporal compositing and statistical characterization, and (iv) exploratory pattern analysis using unsupervised learning.

\subsection{Study Area}
Guayas Province is one of the most densely populated and economically active regions in Ecuador, hosting major urban centers such as Guayaquil, Durán, and Samborondón cantons. According to official population projections, the province concentrates a large share of national industrial, commercial, and transportation activities, making it a relevant case study for satellite-based urban air-quality assessment. The administrative division into cantons provides a natural spatial unit for aggregation and comparison of $NO_2$ pollution patterns.

\subsection{Dataset}



The dataset acquisition and preprocessing stage is based on satellite observations from the Sentinel-5P mission, accessed through the Google Earth Engine (GEE) platform. First, the official Level-3 Offline (OFFL) Sentinel-5P image collection for the selected pollutant (e.g., tropospheric $NO_2$ column density) is loaded and spatially filtered to the administrative boundaries of Guayas province, Ecuador, using FAO GAUL level-1 and level-2 geometries. The temporal domain is restricted to the period 2020–2025, and only the relevant pollutant band is selected. As a preprocessing step, a quality-oriented temporal filtering is applied at the monthly scale: for each year, monthly composites (computed using the median to ensure robustness) are aggregated over the province, and months whose median values fall outside de median $\pm$ 2 $\cdot$ (standard deviation) range are identified as outliers and removed, provided that a minimum number of valid months remains. This procedure reduces the influence of anomalous events and noisy observations while preserving interannual representativeness.

\subsection{Statistical Characterization}
Instead of relying on mean values, which can be strongly influenced by clean background areas or isolated outliers, the methodology emphasizes robust distributional metrics derived from the empirical pixel-value distribution within each canton and year. Five summary statistics are considered: the median ($\tilde{x}$), and the 90th, 95th, and 99th percentiles ($P_{90}$, $P_{95}$, and $P_{99}$).

The median represents a typical background pollution level, while upper-tail percentiles characterize increasingly severe pollution conditions affecting a smaller fraction of the urban area. In particular, $P_{90}$ captures conditions experienced by the most polluted decile of pixels, $P_{95}$ highlights more persistent extreme values, and $P_{99}$ emphasizes severe hotspots while remaining more stable than the absolute maximum.

Let a sample of pixel values be $x_1, x_2, \dots, x_n$, and let the order statistics be
\[
x_{(1)} \le x_{(2)} \le \dots \le x_{(n)}.
\]

\textbf{Median.}
\[
\tilde{x}=
\begin{cases}
x_{(\frac{n+1}{2})}, & \text{if } n \text{ is odd},\\[6pt]
\dfrac{x_{(\frac{n}{2})}+x_{(\frac{n}{2}+1)}}{2}, & \text{if } n \text{ is even}.
\end{cases}
\]

\textbf{Percentile (discrete definition).} For $p \in (0,100)$,
\[
P_p \;=\; x_{\left(\left\lceil \frac{p}{100}\,n \right\rceil\right)}.
\]

\textbf{Percentile (interpolated definition).} Define
\[
h=\frac{p}{100}(n+1),\qquad j=\lfloor h\rfloor,\qquad \gamma=h-j.
\]
Then (with endpoint handling),
\[
P_p=
\begin{cases}
x_{(1)}, & h\le 1,\\[4pt]
(1-\gamma)\,x_{(j)}+\gamma\,x_{(j+1)}, & 1<h<n,\\[6pt]
x_{(n)}, & h\ge n.
\end{cases}
\]

\textbf{Specific upper-tail percentiles.} Using the interpolated definition:
\[
h_{90}=\frac{90}{100}(n+1),\quad j_{90}=\lfloor h_{90}\rfloor,\quad \gamma_{90}=h_{90}-j_{90},
\]
\[
P_{90}=(1-\gamma_{90})\,x_{(j_{90})}+\gamma_{90}\,x_{(j_{90}+1)}.
\]

\[
h_{95}=\frac{95}{100}(n+1),\quad j_{95}=\lfloor h_{95}\rfloor,\quad \gamma_{95}=h_{95}-j_{95},
\]
\[
P_{95}=(1-\gamma_{95})\,x_{(j_{95})}+\gamma_{95}\,x_{(j_{95}+1)}.
\]

\[
h_{99}=\frac{99}{100}(n+1),\quad j_{99}=\lfloor h_{99}\rfloor,\quad \gamma_{99}=h_{99}-j_{99},
\]
\[
P_{99}=(1-\gamma_{99})\,x_{(j_{99})}+\gamma_{99}\,x_{(j_{99}+1)}.
\]

By jointly analyzing these metrics, the methodology distinguishes between broad urban background pollution and localized high-emission areas, enabling a nuanced interpretation of spatial heterogeneity and temporal evolution.

\subsection{Temporal and Spatial Analysis}
For each canton, the selected metrics are computed annually, producing multi-year time series that allow the assessment of trends, interannual variability, and potential regime shifts. Comparisons across cantons highlight spatial contrasts within the province and support the identification of consistently high-pollution areas versus more stable or cleaner regions.

To further explore spatial structure, unsupervised clustering (K-means) is applied to the multi-metric feature space composed of median and upper-tail percentiles. This step groups cantons or pixels with similar pollution behavior, facilitating the identification of characteristic pollution regimes without imposing predefined thresholds. The resulting cluster maps provide an intuitive spatial summary of urban $NO_2$ patterns across Guayas Province.

Overall, the proposed methodology offers a statistically robust and interpretable framework for tracking urban atmospheric pollution using Sentinel-5P data, particularly suited to data-scarce regions where ground-based monitoring is limited or unevenly distributed.

\section{\uppercase{Results}}
\label{sec:ExpRes}
This section presents the experimental results obtained by applying the proposed methodology to Sentinel-5P/TROPOMI $NO_2$ data over Guayas Province. Results are organized into quantitative and qualitative analyses, focusing on temporal evolution, spatial contrasts among cantons, and the identification of characteristic pollution patterns.

\begin{figure*}[!h]
\setlength\tabcolsep{0.75pt}
\centering
\scalebox{1.0}{
\begin{tabular}{cc}

$NO_2$ $\tilde{x}$ & $NO_2$ $P_{90}$  \\
\includegraphics[width=.50\textwidth]{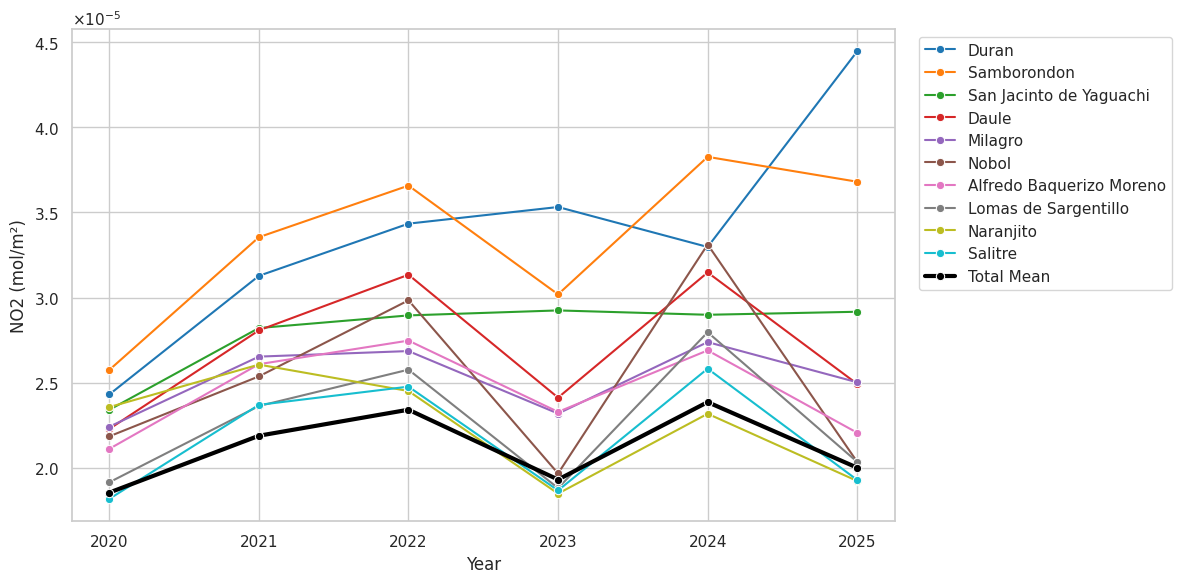} & 
\includegraphics[width=.50\textwidth]{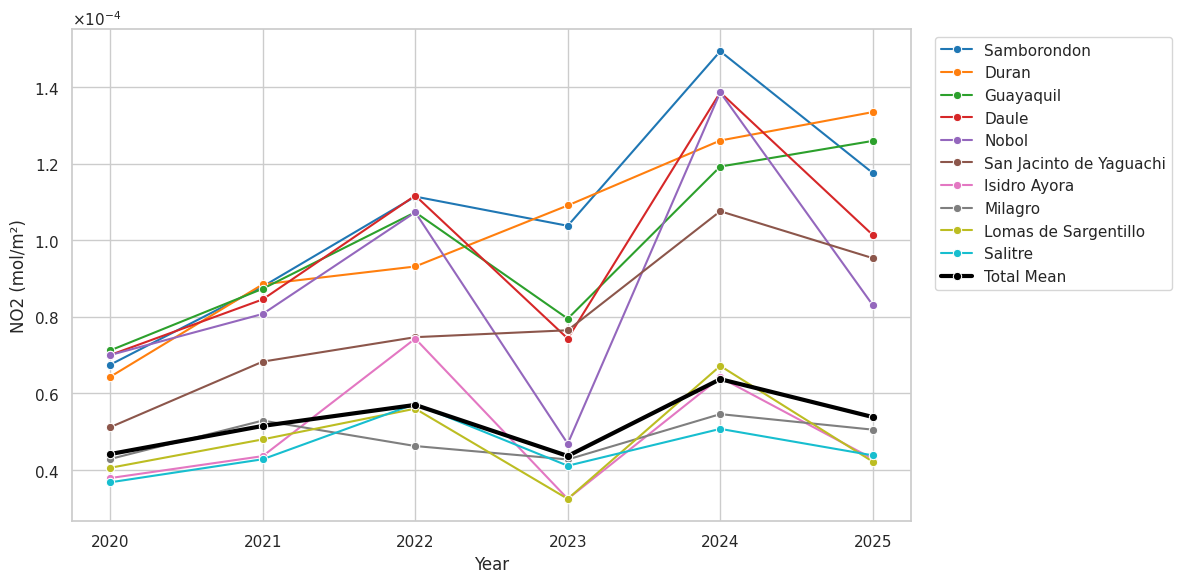} \\

$NO_2$ $P_{95}$ & $NO_2$ $P_{99}$ \\
\includegraphics[width=.50\textwidth]{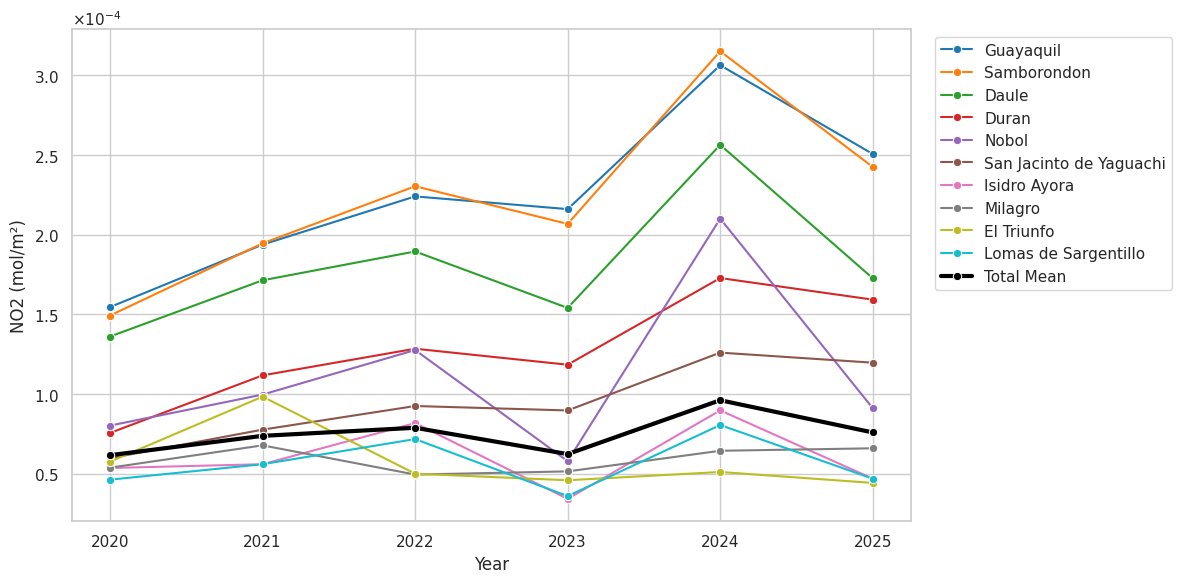} &
\includegraphics[width=.50\textwidth]{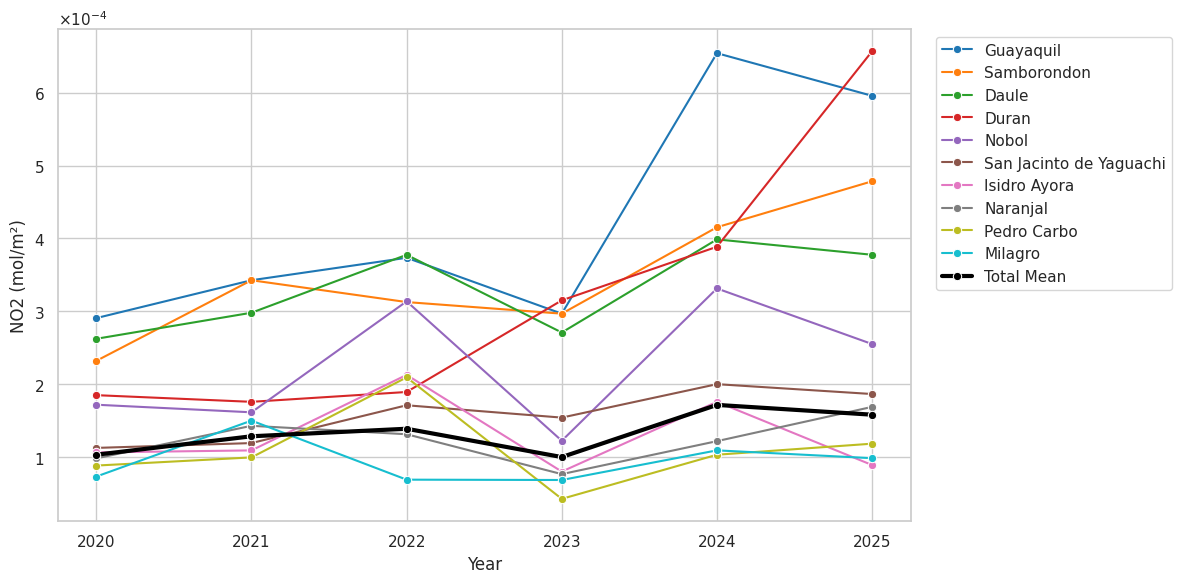}\ \\

\end{tabular}
}
\caption{Spatial distribution of annual $NO_2$ median ($\tilde{x}$), $P_{90}$, $P_{95}$, and $P_{99}$ tropospheric column values over Guayas Province, illustrating background levels and localized pollution hotspots.}
\label{fig:no2_lineplot}
\end{figure*}





\subsection{Quantitative Evaluation}
The quantitative evaluation is based on the annual distributional metrics described in Section \ref{sec:method}, computed for each canton over the study period. The cantons shown in the Table \ref{tab:top10canton} are based on the number of inhabitants according to official data\footnote{https://www.ecuadorencifras.gob.ec/proyecciones-poblacionales/}. Table \ref{tab:top10canton} summarizes the $\tilde{x}$ and upper-tail percentiles ($P_{90}$, $P_{95}$, and $P_{99}$) of tropospheric $NO_2$ columns for the most representative cantons in Guayas Province.

Across all cantons, the $\tilde{x}$ values exhibit relatively moderate interannual variability compared to upper-tail metrics, indicating that background $NO_2$ levels remain comparatively stable over time. In contrast, $P_{90}$, $P_{95}$, and especially $P_{99}$ show substantially larger fluctuations, reflecting the sensitivity of extreme pollution conditions to changes in anthropogenic activity, meteorology, and episodic events.

Highly urbanized and industrialized cantons such as Guayaquil, Durán, and Samborondón consistently present the highest upper-tail values. For example, Guayaquil shows pronounced increases in $P_{95}$ and $P_{99}$ in recent years, suggesting the persistence and, in some cases, intensification of localized $NO_2$ hotspots even when median levels remain relatively stable. This divergence between central and extreme metrics highlights the importance of analyzing the full distribution rather than relying on a single summary statistic.

Less densely populated cantons, including Salitre, El Triunfo, and Empalme, exhibit lower median and percentile values overall, as well as reduced variability in extreme metrics. These patterns are consistent with lower traffic density and industrial activity, supporting the physical interpretability of the satellite-derived indicators.

\subsection{Qualitative Evaluation}
The qualitative evaluation focuses on the spatial distribution of $NO_2$ pollution patterns and their evolution across the study area. Figure \ref{fig:no2_lineplot} presents spatial maps of the median, $P_{90}$, $P_{95}$, and $P_{99}$ $NO_2$ columns, revealing clear spatial gradients and localized hotspots that are not apparent when using average values alone.

Median maps emphasize broad urban–rural contrasts, with elevated background $NO_2$ levels over major urban centers. In contrast, $P_{95}$ and $P_{99}$ maps highlight compact, spatially coherent hotspots associated with dense traffic corridors, industrial zones, and port-related activities. These results confirm that extreme percentiles are effective for isolating persistent high-emission areas within heterogeneous urban environments.

\begin{figure*}[!h]
\setlength\tabcolsep{0.75pt}
\centering
\scalebox{1.0}{
\begin{tabular}{cccccccc}

& 2020 & 2021 & 2022 & 2023 & 2024 & 2025 \\
& & & & & & \\
\rotatebox{90}{$NO_2$ $\tilde{x}$} &
\includegraphics[width=.145\textwidth]{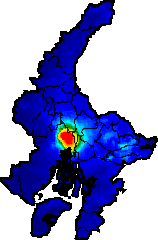} & 
\includegraphics[width=.145\textwidth]{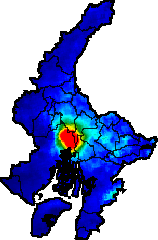} & 
\includegraphics[width=.145\textwidth]{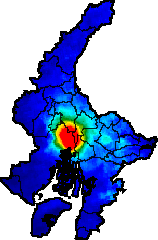} &
\includegraphics[width=.145\textwidth]{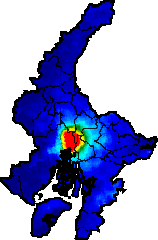} &
\includegraphics[width=.145\textwidth]{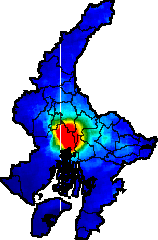} &
\includegraphics[width=.145\textwidth]{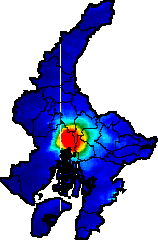} &
\includegraphics[width=.07\textwidth,height=3.5cm]{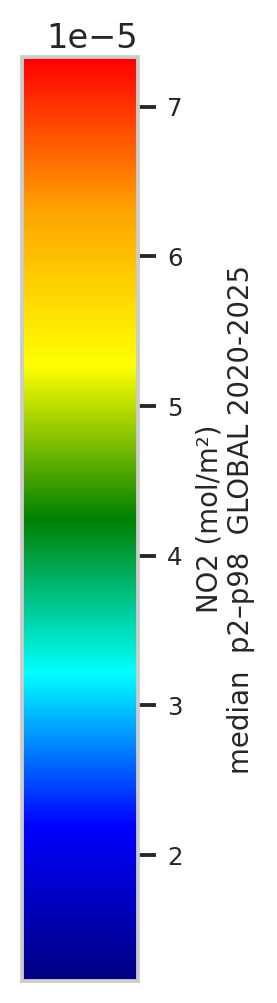} 
\\

\rotatebox{90}{$NO_2$ $P_{90}$} &
\includegraphics[width=.145\textwidth]{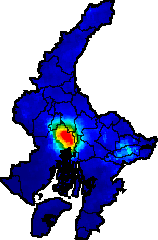} & 
\includegraphics[width=.145\textwidth]{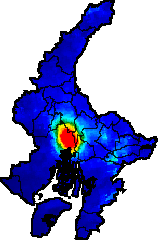} & 
\includegraphics[width=.145\textwidth]{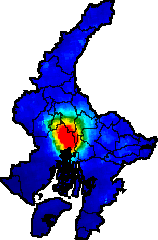} &
\includegraphics[width=.145\textwidth]{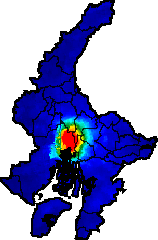} &
\includegraphics[width=.145\textwidth]{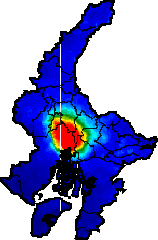} &
\includegraphics[width=.145\textwidth]{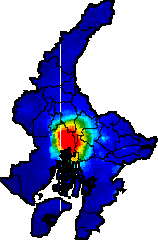} &
\includegraphics[width=.07\textwidth,height=3.5cm]{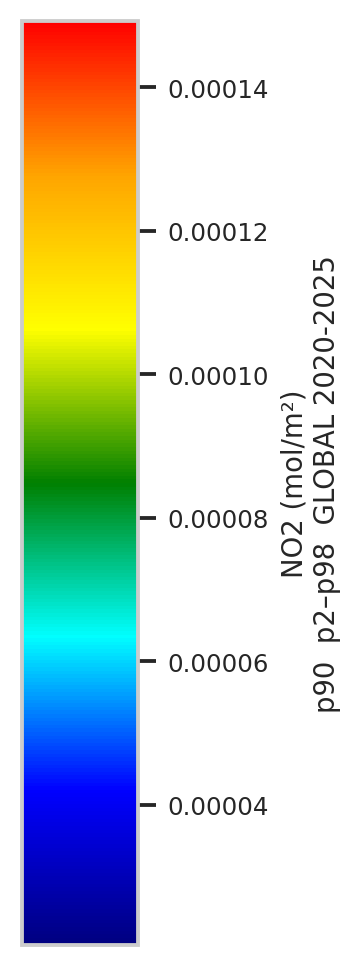} 
\\

\rotatebox{90}{$NO_2$ $P_{95}$} &
\includegraphics[width=.145\textwidth]{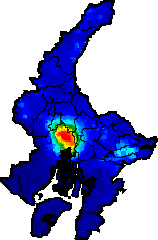} & 
\includegraphics[width=.145\textwidth]{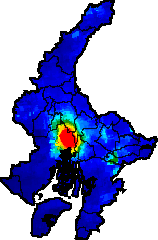} & 
\includegraphics[width=.145\textwidth]{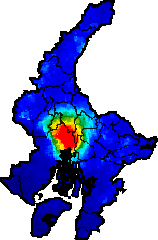} &
\includegraphics[width=.145\textwidth]{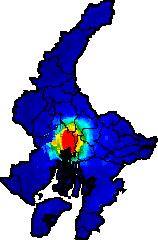} &
\includegraphics[width=.145\textwidth]{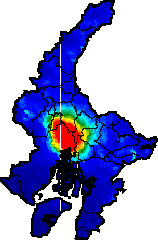} &
\includegraphics[width=.145\textwidth]{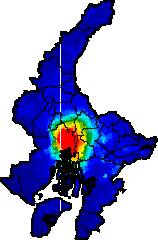} &
\includegraphics[width=.07\textwidth,height=3.5cm]{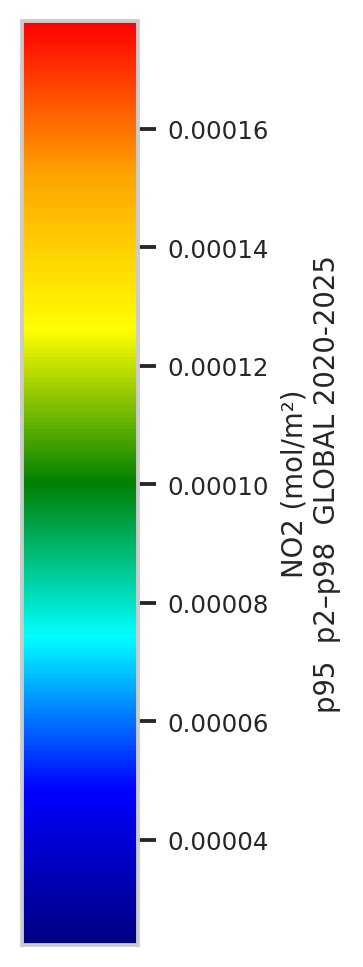} 
\\

\rotatebox{90}{$NO_2$ $P_{99}$} &
\includegraphics[width=.145\textwidth]{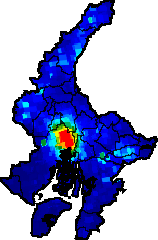} & 
\includegraphics[width=.145\textwidth]{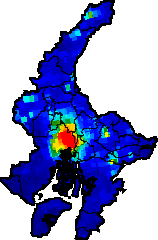} & 
\includegraphics[width=.145\textwidth]{images/no2/no2_p95_raster_2022.png} &
\includegraphics[width=.145\textwidth]{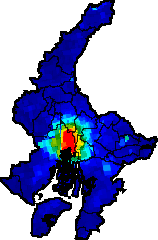} &
\includegraphics[width=.145\textwidth]{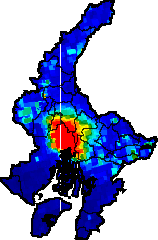} &
\includegraphics[width=.145\textwidth]{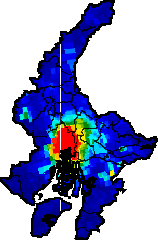} &
\includegraphics[width=.07\textwidth,height=3.5cm]{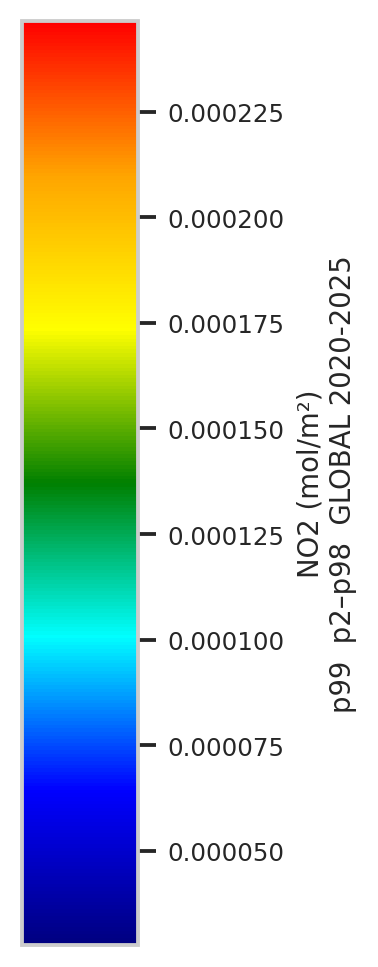} 
\\
& & & & & & \\
\end{tabular}
}
\caption{Interannual evolution (2020–2025) of provincial-scale $NO_2$ distributional metrics ($\tilde{x}$, $P_{90}$, $P_{95}$, and $P_{99}$), highlighting the stronger variability of extreme percentiles relative to background conditions.}
\label{fig:no2_maps}
\end{figure*}

Figure \ref{fig:no2_maps} illustrates the temporal evolution of the median and upper-tail percentiles aggregated at the provincial level. The figure shows that extreme percentiles respond more strongly to interannual changes than the median, reinforcing their utility for tracking changes in urban pollution severity.

To further synthesize spatial patterns, an intensity-based K-means clustering is applied to canton-level $NO_2$ statistics. For each canton, three features are extracted from the annual $NO_2$ distributions: the mean, maximum, and standard deviation of the selected $NO_2$ metric across the study period. Prior to clustering, features are standardized using z-score normalization to ensure equal contribution of all variables.

\begin{figure*}[!h]
\setlength\tabcolsep{0.75pt}
\centering
\scalebox{1.0}{
\begin{tabular}{cccc}

$NO_2$ $\tilde{x}$ & $NO_2$ $P_{90}$ & $NO_2$ $P_{95}$ & $NO_2$ $P_{99}$ \\
\includegraphics[width=.25\textwidth]{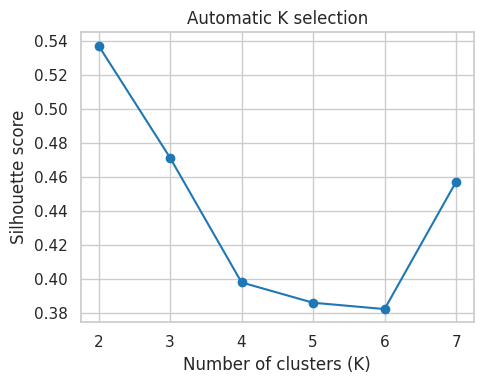} & 
\includegraphics[width=.25\textwidth]{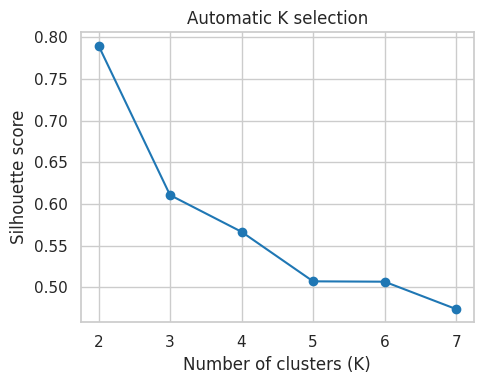} &
\includegraphics[width=.25\textwidth]{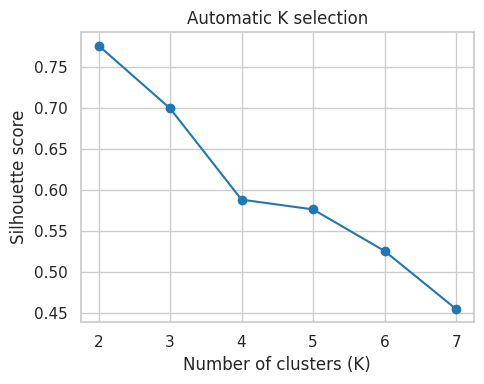} &
\includegraphics[width=.25\textwidth]{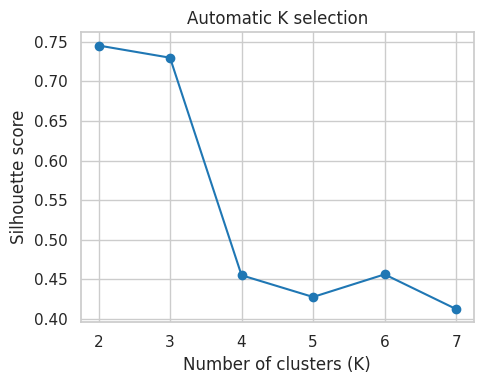}\ \\

\end{tabular}
}
\caption{Silhouette-based automatic K selection for K-means clustering using $NO_2$ distributional metrics, showing the optimal number of clusters (K = 2) across median and upper-tail percentiles.}
\label{fig:no2_maps_elbowing}
\end{figure*}

The optimal number of clusters (K) is determined automatically using the silhouette score, evaluated for values of K ranging from 2 to 7 (see Fig. \ref{fig:no2_maps_elbowing}). The silhouette analysis consistently indicated an optimal solution at $K = 2$, reflecting a clear separation between cantons characterized by higher $NO_2$ intensity and variability and those exhibiting lower and more homogeneous pollution levels.

\begin{figure*}[!h]
\setlength\tabcolsep{0.75pt}
\centering
\scalebox{1.0}{
\begin{tabular}{cccc}

$NO_2$ $\tilde{x}$ & $NO_2$ $P_{90}$ & $NO_2$ $P_{95}$ & $NO_2$ $P_{99}$ \\

\includegraphics[width=.165\textwidth]{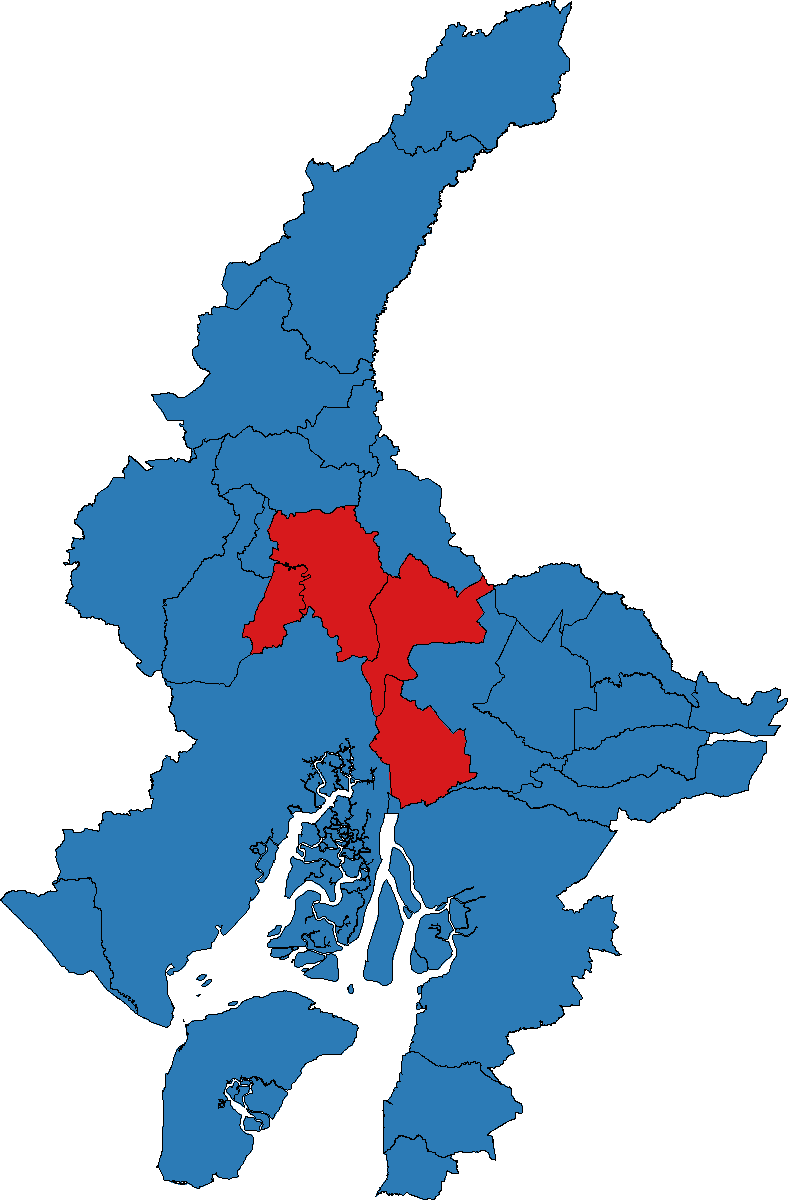} & 
\includegraphics[width=.165\textwidth]{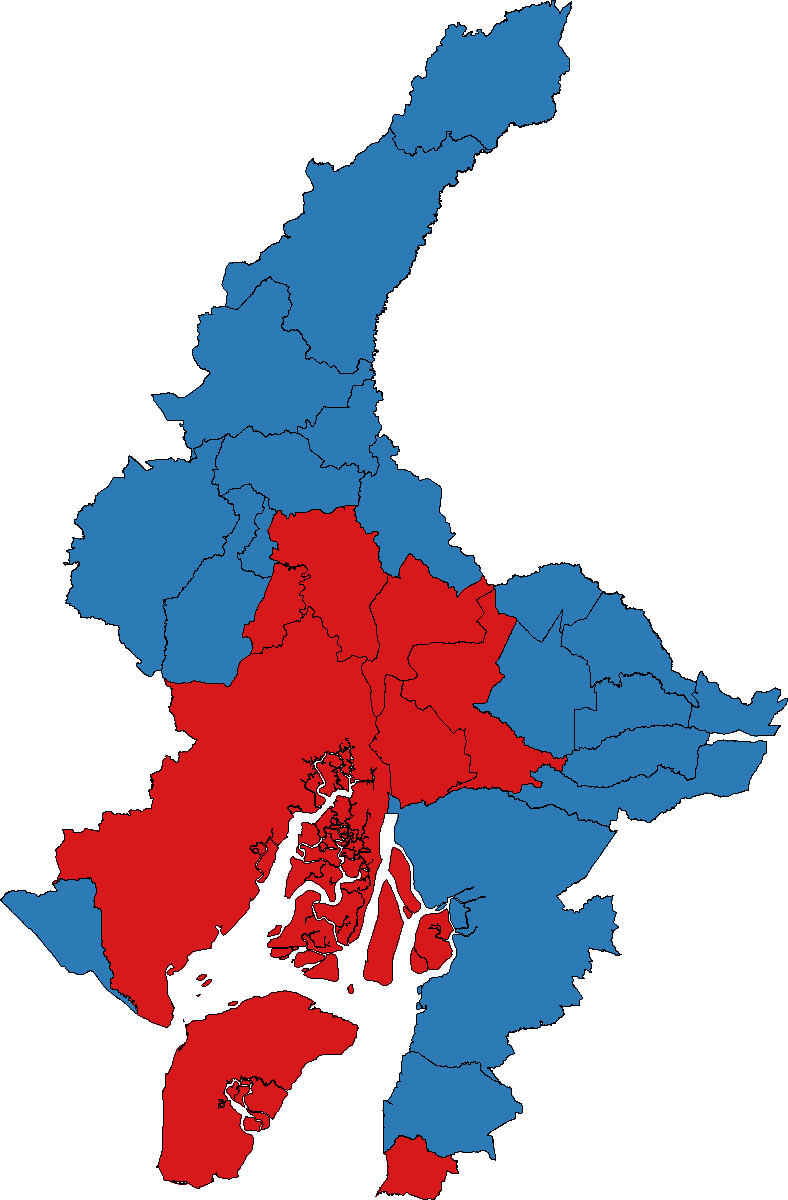} & 
\includegraphics[width=.165\textwidth]{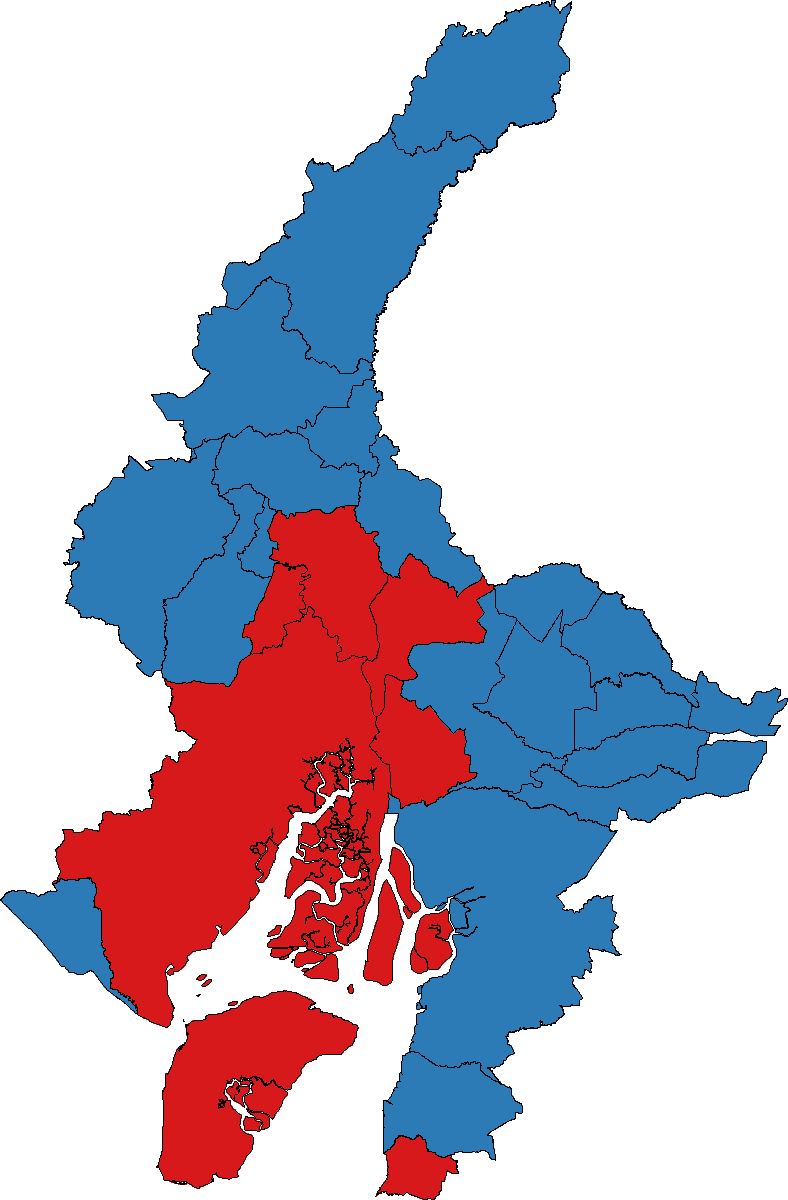} &
\includegraphics[width=.165\textwidth]{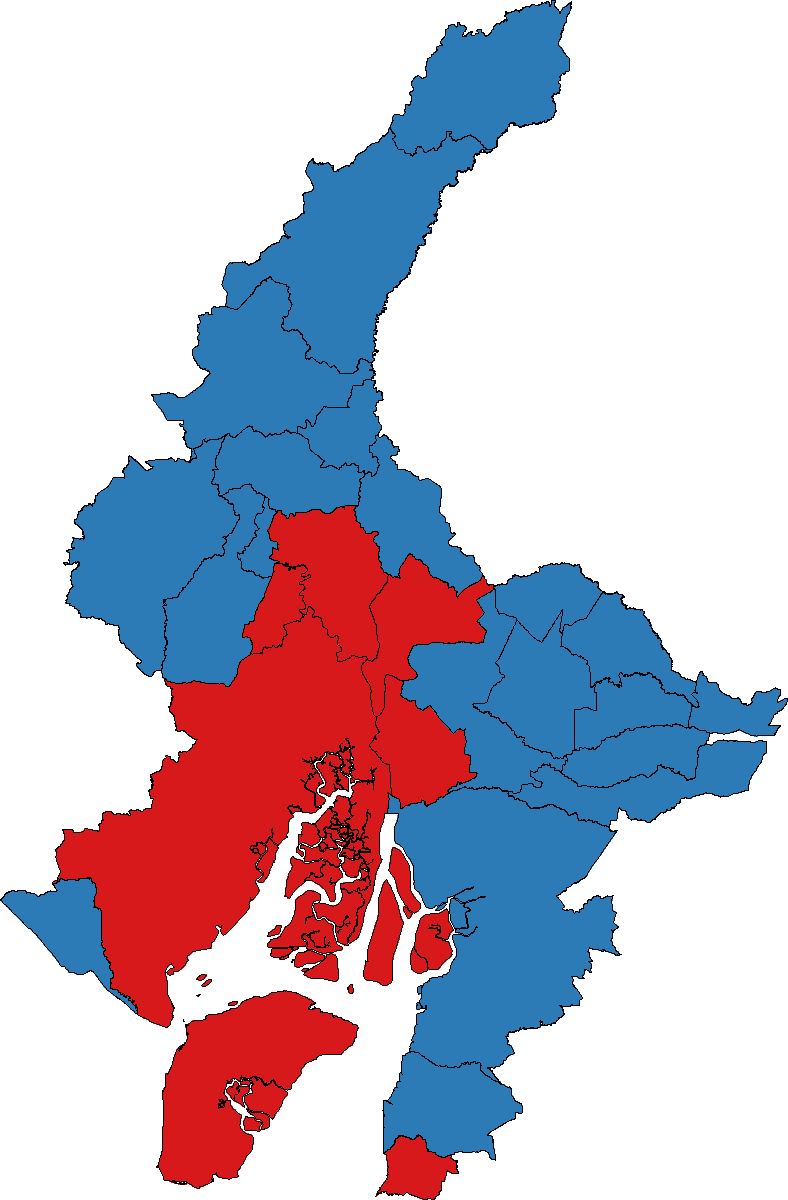} \\

Daule & Daule & Daule & Daule \\
Durán & Durán & Durán & Durán \\
Nobol & Nobol & Nobol & Nobol \\
Samborondon & Samborondon & Samborondon & Samborondon \\
- & San Jacinto de Yaguachi & Guayaquil & Guayaquil \\
- & Guayaquil & - & - \\

\end{tabular}
}
\caption{K-means clustering map of Guayas Province based on $NO_2$ distributional metrics, grouping cantons with similar pollution intensity and variability patterns. The names of the cantons highlighted in red (cluster with high value of $NO_2$) are listed below the map.}
\label{fig:no2_maps_kmeans}
\end{figure*}

Figure \ref{fig:no2_maps_kmeans} shows the K-means clustering results obtained from canton-level $NO_2$ distributional metrics derived from Sentinel-5P/TROPOMI data. The clustering, based on standardized mean, maximum, and standard deviation of $NO_2$ columns, identifies two clearly distinct pollution regimes across Guayas Province.

The cluster highlighted in red corresponds to cantons with higher $NO_2$ intensity and variability, characterized by elevated median values and pronounced upper-tail percentiles. This group is dominated by highly urbanized and industrial cantons, including Guayaquil, Durán, Daule, and Samborondón, where dense traffic networks, port activities, and concentrated economic activity contribute to persistent $NO_2$ hotspots.

In contrast, the remaining cantons form a low-intensity cluster with lower NO$_2$ levels and reduced variability, consistent with less urbanized or predominantly rural environments. The strong spatial coherence of the clusters and their agreement with known urbanization patterns support the interpretability and robustness of the proposed unsupervised classification approach.

Overall, the experimental results demonstrate that the proposed methodology successfully captures both temporal dynamics and spatial heterogeneity of urban $NO_2$ pollution using Sentinel-5P data. The combined use of robust statistical metrics and unsupervised learning provides an interpretable and scalable framework for urban air-quality assessment in data-scarce regions.

\begin{table*}[!h]
    \centering
    \resizebox{2.00\columnwidth}{!}{
    \begin{tabular}{crc|rrrrrr}
         \toprule
         \multirow{2}{*}{Canton} & \multirow{2}{*}{Population} & \multirow{2}{*}{Metric} & \multicolumn{6}{c}{$NO_2\ (\times 10^{-5})$} \\
         & & & 2020 & 2021 & 2022 & 2023 & 2024 & 2025 \\
         \midrule
\multirow{4}{*}{Guayaquil} &
\multirow{4}{*}{2,924,038} 
& $\tilde{x}$ & 1.41 & 1.64 & 1.69 & 1.60 & 1.67 & 1.60 \\
&& $P_{90}$      & 7.12 & 8.73 & 10.73 & 7.95 & 11.92 & 12.59 \\
&& $P_{95}$      & 15.46 & 19.38 & 22.41 & 21.61 & 30.64 & 25.07 \\
&& $P_{99}$      & 29.09 & 34.28 & 37.36 & 29.70 & 65.40 & 59.56 \\

\midrule
\multirow{4}{*}{Durán} &
\multirow{4}{*}{319,622} 
& $\tilde{x}$   & 2.43 & 3.13 & 3.43 & 3.53 & 3.30 & 4.45 \\
&& $P_{90}$      & 6.43 & 8.85 & 9.31 & 10.91 & 12.60 & 13.35 \\
&& $P_{95}$      & 7.56 & 11.18 & 12.85 & 11.85 & 17.28 & 15.93 \\
&& $P_{99}$      & 18.52 & 17.61 & 18.96 & 31.52 & 38.87 & 65.67 \\

\midrule
\multirow{4}{*}{Daule} &
\multirow{4}{*}{228,833} 
& $\tilde{x}$   & 2.23 & 2.81 & 3.13 & 2.41 & 3.15 & 2.49 \\
&& $P_{90}$      & 7.00 & 8.46 & 11.17 & 7.43 & 13.86 & 10.14 \\
&& $P_{95}$      & 13.61 & 17.14 & 18.95 & 15.41 & 25.64 & 17.27 \\
&& $P_{99}$      & 26.25 & 29.82 & 37.76 & 27.11 & 39.88 & 37.77 \\

\midrule
\multirow{4}{*}{Milagro} &
\multirow{4}{*}{212,982} 
& $\tilde{x}$   & 2.24 & 2.65 & 2.69 & 2.32 & 2.74 & 2.50 \\
&& $P_{90}$      & 4.29 & 5.29 & 4.63 & 4.28 & 5.46 & 5.05 \\
&& $P_{95}$      & 5.38 & 6.78 & 4.95 & 5.15 & 6.44 & 6.60 \\
&& $P_{99}$      & 7.37 & 15.04 & 6.94 & 6.89 & 10.96 & 9.89 \\

\midrule
\multirow{4}{*}{Samborondon} &
\multirow{4}{*}{103,266} 
& $\tilde{x}$   & 2.57 & 3.35 & 3.66 & 3.02 & 3.83 & 3.68 \\
&& $P_{90}$      & 6.75 & 8.79 & 11.14 & 10.38 & 14.94 & 11.76 \\
&& $P_{95}$      & 14.93 & 19.46 & 23.05 & 20.68 & 31.52 & 24.26 \\
&& $P_{99}$      & 23.23 & 34.28 & 31.29 & 29.70 & 41.54 & 47.84 \\

\midrule
\multirow{4}{*}{Naranjal} &
\multirow{4}{*}{90,511} 
& $\tilde{x}$   & 1.62 & 2.07 & 2.29 & 2.01 & 2.05 & 1.93 \\
&& $P_{90}$      & 3.47 & 4.61 & 4.00 & 4.00 & 3.95 & 4.74 \\
&& $P_{95}$      & 4.67 & 6.34 & 4.89 & 5.21 & 4.99 & 6.63 \\
&& $P_{99}$      & 9.86 & 14.32 & 13.17 & 7.70 & 12.22 & 16.92 \\

\midrule
\multirow{4}{*}{Empalme} &
\multirow{4}{*}{87,296} 
& $\tilde{x}$   & 1.50 & 1.73 & 1.96 & 1.65 & 2.02 & 1.52 \\
&& $P_{90}$      & 3.36 & 4.23 & 3.94 & 2.97 & 3.98 & 3.36 \\
&& $P_{95}$      & 4.29 & 4.72 & 4.79 & 3.61 & 5.21 & 4.11 \\
&& $P_{99}$      & 8.57 & 11.31 & 6.55 & 8.13 & 13.02 & 5.13 \\

\midrule
\multirow{4}{*}{San Jacinto de Yaguachi} &
\multirow{4}{*}{76,761} 
& $\tilde{x}$   & 2.34 & 2.82 & 2.89 & 2.92 & 2.90 & 2.92 \\
&& $P_{90}$      & 5.12 & 6.83 & 7.47 & 7.65 & 10.76 & 9.53 \\
&& $P_{95}$      & 5.98 & 7.77 & 9.25 & 8.97 & 12.60 & 11.97 \\
&& $P_{99}$      & 11.29 & 11.96 & 17.14 & 15.45 & 20.03 & 18.68 \\

\midrule
\multirow{4}{*}{Salitre} &
\multirow{4}{*}{65,470} 
& $\tilde{x}$   & 1.82 & 2.37 & 2.48 & 1.87 & 2.58 & 1.93 \\
&& $P_{90}$      & 3.68 & 4.28 & 5.73 & 4.11 & 5.07 & 4.38 \\
&& $P_{95}$      & 4.24 & 4.87 & 7.16 & 4.76 & 6.19 & 5.89 \\
&& $P_{99}$      & 5.22 & 6.52 & 9.92 & 5.47 & 9.41 & 10.20 \\

\midrule
\multirow{4}{*}{El Triunfo} & 
\multirow{4}{*}{63,924} 
& $\tilde{x}$   & 1.99 & 2.16 & 2.42 & 1.93 & 2.18 & 1.81 \\
&& $P_{90}$      & 4.68 & 5.34 & 4.35 & 3.37 & 4.11 & 3.55 \\
&& $P_{95}$      & 5.76 & 9.84 & 5.01 & 4.59 & 5.11 & 4.43 \\
&& $P_{99}$      & 8.21 & 14.32 & 8.10 & 5.93 & 11.13 & 8.25 \\

         \bottomrule
    \end{tabular}
    }
    \caption{Annual median ($\tilde{x}$) and upper-tail percentiles ($P_{90}$, $P_{95}$, $P_{99}$) of tropospheric $NO_2$ columns derived from Sentinel-5P/TROPOMI for the ten most populated cantons of Guayas Province during the period 2020–2025. Cantons are ordered by population size. These metrics summarize background conditions (median) and increasingly severe pollution extremes ($P_{90}$, $P_{95}$, and $P_{99}$).
    }
    \label{tab:top10canton}
\end{table*}








\section{\uppercase{Conclusions}}
\label{sec:conclusion}
This study shows that Sentinel-5P/TROPOMI $NO_2$ data, combined with robust statistics and unsupervised machine learning, can effectively describe urban pollution patterns in regions with scarce monitoring. Instead of relying on averages, it uses distribution metrics (e.g., percentiles) to capture both typical conditions and extreme $NO_2$ events. In Guayas Province, the most urbanized cantons present consistently higher upper-tail $NO_2$ percentiles even when median values are stable, suggesting persistent localized emission hotspots. K-means clustering based on intensity separates the area into two distinct pollution regimes that align with known differences in urbanization, traffic, and economic activity. Overall, the framework is scalable and reproducible for satellite-based urban air-quality assessment. Although it cannot directly estimate surface concentrations, focusing on relative contrasts, extremes, and spatiotemporal variability makes it useful for long-term monitoring and comparisons where ground data are limited.

\bibliographystyle{apalike}
{\small
\bibliography{example.bib}}

@article{sha2021validation,
  title={Validation of methane and carbon monoxide from Sentinel-5 Precursor using TCCON and NDACC-IRWG stations},
  author={Sha, Mahesh Kumar and Langerock, Bavo and Blavier, Jean-Fran{\c{c}}ois L and Blumenstock, Thomas and Borsdorff, Tobias and Buschmann, Matthias and Dehn, Angelika and De Mazi{\`e}re, Martine and Deutscher, Nicholas M and Feist, Dietrich G and others},
  journal={Atmospheric Measurement Techniques Discussions},
  volume={2021},
  pages={1--84},
  year={2021},
  publisher={G{\"o}ttingen, Germany}
}

@article{lorente2021methane,
  title={Methane retrieved from TROPOMI: improvement of the data product and validation of the first 2 years of measurements},
  author={Lorente, Alba and Borsdorff, Tobias and Butz, Andre and Hasekamp, Otto and aan de Brugh, Joost and Schneider, Andreas and Wu, Lianghai and Hase, Frank and Kivi, Rigel and Wunch, Debra and others},
  journal={Atmospheric Measurement Techniques},
  volume={14},
  number={1},
  pages={665--684},
  year={2021},
  publisher={Copernicus Publications G{\"o}ttingen, Germany}
}

@article{tian2022investigating,
  title={Investigating the performance of carbon monoxide and methane observations from Sentinel-5 precursor in China},
  author={Tian, Yuan and Hong, Xinhua and Shan, Changgong and Sun, Youwen and Wang, Wei and Zhou, Minqiang and Wang, Pucai and Lin, Peize and Liu, Cheng},
  journal={Remote Sensing},
  volume={14},
  number={23},
  pages={6045},
  year={2022},
  publisher={MDPI}
}

@article{fenger1999urban,
  title={Urban air quality},
  author={Fenger, Jes},
  journal={Atmospheric environment},
  volume={33},
  number={29},
  pages={4877--4900},
  year={1999},
  publisher={Elsevier}
}

@article{andreae2019emission,
  title={Emission of trace gases and aerosols from biomass burning--an updated assessment},
  author={Andreae, Meinrat O},
  journal={Atmospheric Chemistry and Physics},
  volume={19},
  number={13},
  pages={8523--8546},
  year={2019},
  publisher={Copernicus Publications G{\"o}ttingen, Germany}
}

@article{Douros2023GMD,
  author  = {Douros, John and Eskes, Henk and van Geffen, Jos and Boersma, K. Folkert and Compernolle, Steven and Pinardi, Gaia and Blechschmidt, Anne-Marlene and Peuch, Vincent-Henri and Colette, Augustin and Veefkind, J. Pepijn},
  title   = {Comparing Sentinel-5P TROPOMI NO$_2$ column observations with the CAMS regional air quality ensemble},
  journal = {Geoscientific Model Development},
  year    = {2023},
  volume  = {16},
  pages   = {509--534},
  doi     = {10.5194/gmd-16-509-2023},
  url     = {https://gmd.copernicus.org/articles/16/509/2023/}
}

@article{Judd2020AMT,
  author  = {Judd, Laura M. and Al-Saadi, Jassim A. and Szykman, James J. and Valin, Lukas C. and Janz, Scott J. and Kowalewski, Matthew G. and Eskes, Henk J. and Veefkind, J. Pepijn and Cede, Alexander and others},
  title   = {Evaluating Sentinel-5P TROPOMI tropospheric NO$_2$ column densities with airborne and Pandora spectrometers near New York City and Long Island Sound},
  journal = {Atmospheric Measurement Techniques},
  year    = {2020},
  volume  = {13},
  number  = {11},
  pages   = {6113--6140},
  doi     = {10.5194/amt-13-6113-2020},
  url     = {https://pmc.ncbi.nlm.nih.gov/articles/PMC8193800/}
}

@article{Verhoelst2021AMT,
  author  = {Verhoelst, Tijl and Compernolle, Steven and Pinardi, Gaia and Lambert, Jean-Christopher and Eskes, Henk J. and Eichmann, Kai-Uwe and Fj{\ae}raa, Ann Mari and Granville, Jos{\'e} and Niemeijer, Sander and Cede, Alexander and others},
  title   = {Ground-based validation of the Copernicus Sentinel-5P TROPOMI NO$_2$ measurements with the NDACC ZSL-DOAS, MAX-DOAS and Pandonia global networks},
  journal = {Atmospheric Measurement Techniques},
  year    = {2021},
  volume  = {14},
  pages   = {481--510},
  doi     = {10.5194/amt-14-481-2021},
  url     = {https://amt.copernicus.org/articles/14/481/2021/}
}

@article{mejia2023sentinel,
  title={Sentinel satellite data monitoring of air pollutants with interpolation methods in Guayaquil, Ecuador},
  author={Mej{\'\i}a, Danilo and Alvarez, Hermel and Zalakeviciute, Rasa and Macancela, Diana and Sanchez, Carlos and Bonilla, Santiago},
  journal={Remote Sensing Applications: Society and Environment},
  volume={31},
  pages={100990},
  year={2023},
  publisher={Elsevier}
}

@inproceedings{davybida2023air,
  title={Air quality impacts of war detected from the Sentinel-5P satellite over Ukraine},
  author={Davybida, LI},
  booktitle={Conf. Series: Earth and Environmental Science},
  volume={1254},
  number={1},
  pages={012112},
  year={2023},
  organization={IOP Publishing}
}

@article{shah2024air,
  title={Air Quality Monitoring Using Sentinel-5p TROPOMI—A Case Study of Pune City},
  author={Shah, Suraj V and Gaikwad, Sandeep V and Vibhute, Amol D},
  journal={SN Computer Science},
  volume={5},
  number={8},
  pages={1125},
  year={2024},
  publisher={Springer}
}

@article{fiore2015air,
  title={Air quality and climate connections},
  author={Fiore, Arlene M and Naik, Vaishali and Leibensperger, Eric M},
  journal={Journal of the Air \& Waste Management Association},
  volume={65},
  number={6},
  pages={645--685},
  year={2015},
  publisher={Taylor \& Francis}
}

@article{prajesh2025satellite,
  title={Satellite-Based Seasonal Fingerprinting of Methane Emissions from Canadian Dairy Farms Using Sentinel-5P},
  author={Prajesh, Padmanabhan Jagannathan and Ragunath, Kaliaperumal and Gordon, Miriam and Neethirajan, Suresh},
  journal={Climate},
  volume={13},
  number={7},
  pages={135},
  year={2025},
  publisher={MDPI}
}

@inproceedings{velesaca2024analysis,
  title={Analysis of Hidden Patterns in Road Accident Dataset Using Clustering Techniques},
  author={Velesaca, Henry O and Realpe, Miguel and Sappa, Angel D and Gomez, Alice},
  booktitle={International Conference on Smart Technologies, Systems and Applications},
  pages={96--111},
  year={2024},
  organization={Springer}
}







\end{document}